\documentclass[final,5p,times,12pt,letter]{elsarticle}

 \usepackage{graphicx}

\usepackage{amssymb}

\journal{Science \& Public Policy (SPPI)}

\begin{document}

\begin{frontmatter}



\title{{\huge \textsc{Climate Change and Its Causes}} \\
 A Discussion About Some Key Issues}


\author{Nicola Scafetta $^{1,2}$
}

 \address{$^{1}$Active Cavity Radiometer Irradiance Monitor (ACRIM) Lab, Coronado, CA 92118, USA}

 \address{$^{2}$Department of Physics, Duke University, Durham, NC 27708, USA.}


\begin{abstract}
This article discusses the limits of the Anthropogenic Global Warming Theory  advocated by the Intergovernmental Panel on Climate Change. A phenomenological theory
of climate change based on the physical properties of the data themselves is proposed. At least 60\% of the warming of the Earth observed since 1970 appears to be induced by
natural cycles which are present in the solar system. A climatic stabilization or cooling until 2030-2040 is forecast by the phenomenological model. \newline
\\ ****************************************************************************************\newline
\\
{\small This work is made of

\begin{itemize}
  \item An  translation into English of the paper:  \\Scafetta N., ``Climate Change and Its causes: A Discussion about Some Key Issues,'' \emph{La Chimica e l'Industria} \textbf{1}, p. 70-75 (2010);
  \item Several additional supporting notes are added to the paper;
  \item An extended appendix section part is added to cover several thematic issues to support particular topics addressed in the main paper. (not in this preprint)
\end{itemize}

This work covers most topics presented by Scafetta at a seminar at the U.S. Environmental Protection Agency, DC USA, February 26, 2009.} A video of the seminar is here: \newline
http://yosemite.epa.gov/ee/epa/eed.nsf/vwpsw/360796B06E48EA0485257601005982A1\#video
\newline \\
 Cite as: \newline  Scafetta N., ``Climate Change and Its causes: A Discussion about Some Key Issues,''
 \emph{La Chimica e l'Industria} \textbf{1}, p. 70-75 (2010).
\newline\\
The full English version with the appendixes can be downloaded from\\
 http://scienceandpublicpolicy.org/originals/climate\_change\_causes.html
\newline\\
The Italian version of the original paper can be downloaded (with possible journal restrictions) from \\http://www.soc.chim.it/files/chimind/pdf/2010/2010\_1\_70.pdf

\end{abstract}

\onecolumn

\end{frontmatter}

\onecolumn

{\large

\section{Introduction}
\begin{figure*}[t]
\centering
\includegraphics[width=12.0cm]{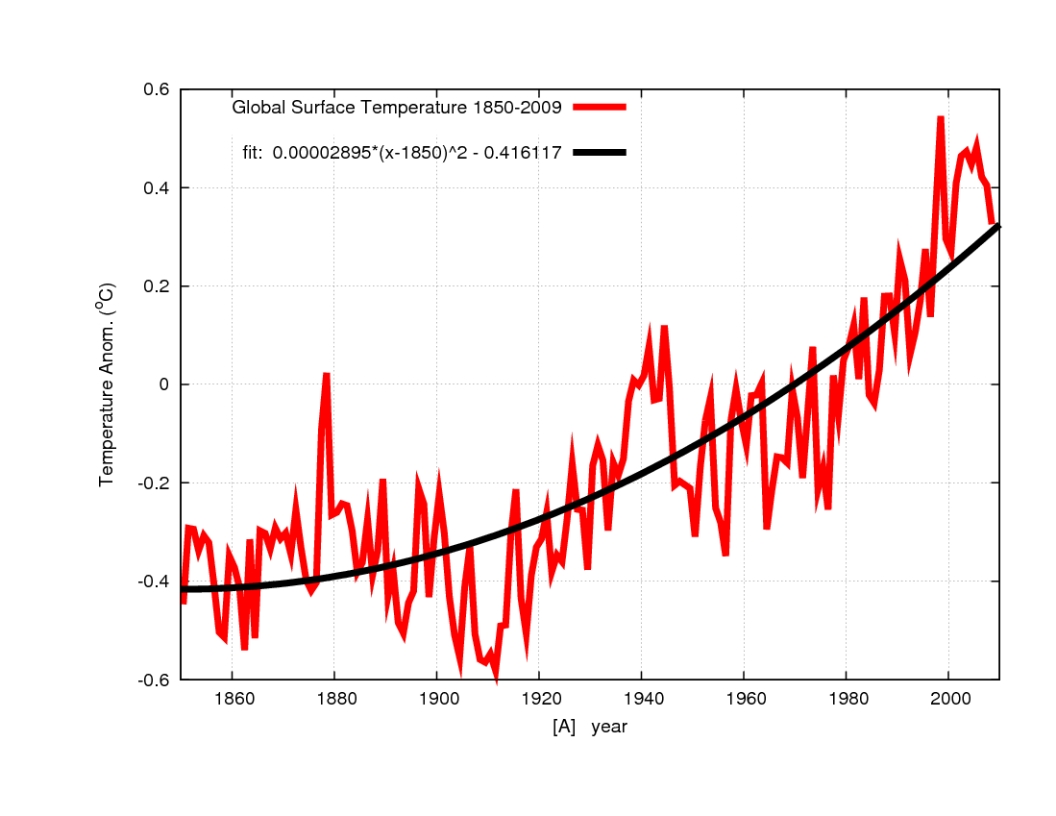}
\caption{
Global surface temperature (land and sea) HadCRUT3 (red) and its quadratic fit (black). [Climatic Research Unit, http://www.cru.uea.ac.uk/].  }
\end{figure*}

Since 1900 the global surface temperature of the Earth has risen by about 0.8 $^oC$ (Figure 1), and since
the 70s  by about 0.5 $^oC$.
This temperature increase occurred during a significant atmospheric concentration increase
of some greenhouse gases, especially $CO_2$ and $CH_4$, which is known to be mainly due to human emissions. According to the \emph{Anthropogenic Global Warming Theory} (AGWT) humans have caused more
 than 90\% of global warming since 1900 and virtually 100\% of the global warming since 1970 (Appendix A). The AGWT is currently advocated by the \emph{Intergovernmental Panel on Climate Change} (IPCC) [1], which is the leading body for the assessment of climate change established by the United Nations Environment Programme (UNEP) and the World Meteorological Organization (WMO). Many scientists believe that further emissions of greenhouse gases could endanger humanity [2].\\

However, not everyone shares the IPCC's views [3].\footnote{The AGWT advocates claim that there exists a  \emph{scientific consensus} that supports the AGWT. However,  a \emph{scientific consensus} does not have any scientific value when it is contradicted by data. It is perfectly legitimate  to discuss the topic of manmade global warming and closely scrutinize the IPCC's claims. Given the extreme complexity of the climate system and the overwhelming evidence that climate has always changed, the AGWT advocates' claim that the science is \emph{settled} is premature in the extreme. }
More than 30,000 scientists in America  (including 9,029 PhDs) have recently signed a petition stating that those claims are extreme,
that the climate system is more complex than what is now known, several mechanisms are not yet included
in the climate models considered by the IPCC and that this issue should be treated with some caution
because  incorrect environmental policies could also cause extensive damage [3]. This article briefly
summarizes some of the reasons,  mostly derived from my own research, why the science behind
the IPCC's claim is questionable.\footnote{On November 19, 2009 a \emph{climategate} story erupted on
the web. This story is seriously undermining the credibility of the AGWT and of its advocates.
 Thousands of e-mails and other documents were disseminated via the internet through the hacking of a
 server used by the Climatic Research Unit (CRU) of the University of East Anglia in Norwich,
 England. These e-mails have been interpreted by some as suggesting serious scientific
  misconduct and even conspiracy by leading climate scientists and IPCC authors who have strongly advocated AGWT.
These emails apparently suggest: 1) manipulation of  temperature data; 2) prevention of a proper scientific disclosure of  data and methodologies; 3) attempts to discredit scientists critical of the AGWT also by means of internet articles such as those at http://www.realclimate.org (several of these realclimate.org articles are quite shallow and suspiciously inaccurate); 4) attempts to bias Wikipedia articles in  favor of the AGWT; 5) and much more seriously, attempts to control which papers  appear in the peer reviewed literature and in the climate assessments in such a way to bias the scientific community in favor of the AGWT. Others, however, believe that the contents of those emails have been maliciously misinterpreted by the so-called \emph{skeptics}. A detailed analysis of these emails can be found   in: 1) J. P. Costella (2010), \emph{Climategate analysis}, SPPI reprint series, (http://scienceandpublicpolicy.org/reprint/climategate\_analysis.html); 2) S. Mosher and T. W. Fuller (2010),  \emph{Climategate: The Crutape Letters}, CreateSpace publisher; 3) See also United States Senate Report
`Consensus' Exposed: The CRU Controversy,  http://epw.senate.gov/public/index.cfm?FuseAction=Files.View\&FileStore\_id=7db3fbd8-f1b4-4fdf-bd15-12b7df1a0b63}\\

\begin{figure*}[tbp]
\centering
\includegraphics[width=12.0cm]{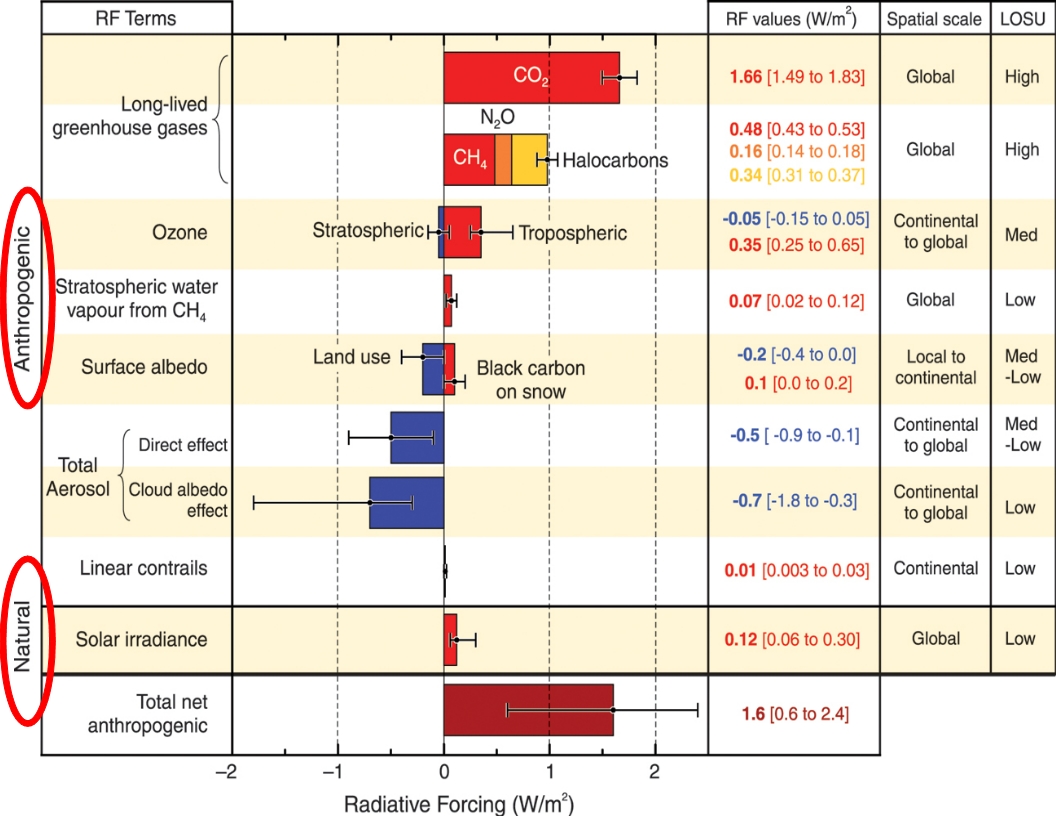}
\caption{
List of radiative forcings held responsible for the global warming since 1750 and used in the models adopted by the IPCC. The figure is adapted from  the \emph{IPCC Climate Change 2007: Synthesis Report}. These forcings are used as inputs of the climate models used by the IPCC to support the AGWT. The table suggests that the total net anthropogenic forcing since 1750 has been 13.3 times larger than the natural forcing. However,  labeling on the left of the table, \emph{anthropogenic} and \emph{natural}, is misleading because it  would imply that only human activity can change the chemistry of the atmosphere, which is non physical.  }
\end{figure*}

\section{The IPCC's pro-anthropogenic warming bias}

First, it should be noted that the IPCC mission states:
\begin{quote}
\emph{``The IPCC reviews and assesses the most recent scientific, technical and socio-economic information produced worldwide relevant to the understanding of human-induced climate change.''}
\end{quote}
The above statement implies that  the IPCC may  provide a colored reading of the scientific literature by stressing those studies that would better justify its own political mission, which evidently focuses on \emph{human-induced} climate change.\footnote{Further evidence of the IPCC's anthropogenic and political  bias was discovered  in January 2010: the IPCC's claim that Himalayan glaciers will disappear by 2035 was based on   magazine interviews, not on peer-reviewed scientific research which is contrary to the IPCC's own policy.   Dr. Lal admitted that this physically impossible event was highlighted  in the IPCC report just to put political pressure on world leaders (http://www.dailymail.co.uk/news/article-1245636/Glacier-scientists-says-knew-data-verified.html). Curiously, NASA anticipated the Himalayan glacier melting to 2030 (http://wattsupwiththat.com/2010/01/23/nasa-climate-page-suckered-by-ipcc-deletes-a-moved-up-glacier-melting-date-reference/).
Other significant errors and non peer-reviewed material in the IPCC have been uncovered such as: Endanger 40 percent of Amazon rainforests; Melt mountain ice in the Alps, Andes, and Africa; Deplete water resources for 4.5 billion people by 2085, neglecting to mention that global warming could also increase water resources for as many as 6 billion people; Slash crop production by 50 percent in North Africa by 2020. http://epw.senate.gov/public/index.cfm?FuseAction=Files.View\&FileStore\_id=9cc0e46e-56be-4728-9099-92dbda199bfc}\\

\begin{figure*}[tbp]
\centering
\includegraphics[width=14.0cm]{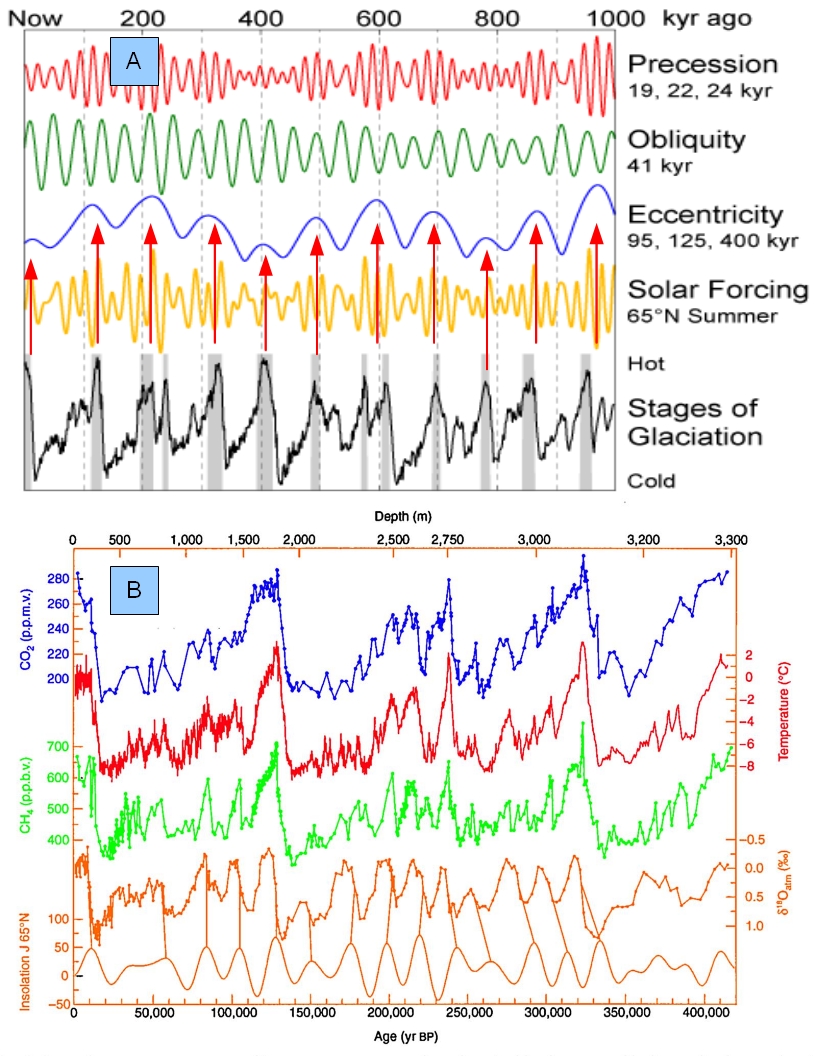}
\caption{
Cycles of $CO_2$ and $CH_4$ that time-lag (with a delay of about 800 years) the temperature cycles observed during the Ice Ages. These glacial cycles were likely induced by the modulation of the solar input into the Earth's system through the Milankovich cycles, which are orbital perturbations of the Earth such as the precession, obliquity and eccentricity. For example, notice the good correspondence between the 100.000 year temperature cycles with the eccentricity, as  highlighted in the figure.  The figure is adapted from Wikipedia' articles: [A] \emph{Milankovitch cycles}; [B] \emph{Vostok Station}. }
\end{figure*}

Indeed, the existence of an anthropogenic bias appears evident in Figure 2 that shows the complete list of the radiative forcings, which, as the IPCC claims, have caused the global climate warming observed since 1750. This figure divides  the climatic forcings  into two groups: one group includes only the total solar irradiance and is labeled \emph{natural}, the other group comprises the rest and  is labeled  \emph{anthropogenic}. Thus, the IPCC claims that 100\% of the increase of the $CO_2$ and $CH_4$ atmospheric concentrations  observed since 1750 and the change of all other climate components, except for the total solar irradiance, are \emph{anthropogenic}. These labels do suggest that without humans the chemical concentrations of the atmosphere and a number of other climatic parameters would remain rigorously unchanged despite a change of the  solar energetic input!\\

This claim is non physical because as the solar activity increases, climate warms, and this causes a net
increase  of atmospheric $CO_2$ and $CH_4$ concentration.  During  warming the ability of the ocean to absorb these gases from the atmosphere decreases because of  Henry's law and other mechanisms. A warming  would also increase  the natural production of atmospheric $CO_2$ and $CH_4$ on the land due to the acceleration of the fermentation of organic material, outgassing of (permafrost) soils  and other mechanisms [3,4]. The existence of $CO_2$ and $CH_4$ feedback mechanisms  are evident in the large $CO_2$ and $CH_4$ cycles  observed during the ice ages (which were caused by the astronomical cycles of Milankovich) when no human industrial activity existed, see Figure 3.\\

For example, even assuming that the IPCC's forcing estimates in Figure 2  are correct, if only 10\% of the total increase in greenhouse gases since 1750 has been due to the observed increase of solar activity during the same period, the IPCC, with its labels, has inflated the anthropogenic contribution by 20\% and  underestimated the solar contribution by 300\%.\footnote{A recent study (using $CO_2$ ice core reconstructions) found a $CO_2$ feedback rate of  1.7-21.4 ppmv $CO_2$ increase per $^oC$, while other theoretical and empirical studies found a larger value  (Frank D. C. \emph{et al.} (2010), Ensemble reconstruction constraints on the global
carbon cycle sensitivity to climate, \emph{Nature} \textbf{463}, 527-530). However, a comparison between global temperature from 1860 and atmospheric $CO_2$ measurements by direct chemical analysis (not by ice core sample reconstruction) shows  an  atmospheric $CO_2$ concentration curve, with a maximum in 1942, which  correlates relatively well  with the  temperature record;  both curves present a maximum in 1940-1945 (Appendix B). This result would indicate the existence of  strong  $CO_2$ feedback mechanisms, which would imply that the observed $CO_2$ concentration increase during the last decades is highly related to some carbon cycle feedback mechanism  in response  to the increased solar input during this same period. It appears that it is climate change that alters  the atmospheric   $CO_2$ concentration, rather  than vice versa. [E.-G. Beck (2007), 180 Years of atmospheric $CO_2$ Gas Analysis by Chemical Methods, \emph{Energy
\& Environment} \textbf{18}, 259-282. http://www.biomind.de/realCO2/realCO2-1.htm ]} This can be easily extrapolated from the numbers depicted in Figure 2.  It is evident that if the climatic forcings are labeled as \emph{anthropogenic},  the presumed consequences, namely  climate changes, would also  be  \emph{anthropogenic}. This, however, is circular logic.\\

\section{The climate sensitivity uncertainty to $CO_2$ increase}

A second fundamental issue  concerns how much global warming can be induced by an increase of  $CO_2$ (or $CH_4$) atmospheric concentration. Indeed, this estimate is extremely uncertain. Also the radiative forcing associated with aerosols is extremely uncertain, as Figure 2 shows.\\

The IPCC acknowledges that if the atmospheric concentration of $CO_2$ doubles the global average temperature could rise between 1.5 and 4.5 $^oC$ at equilibrium.
The variability  of climate sensitivity to $CO_2$ is shown in Figure 4, which demonstrates  an even wider sensitivity temperature range [5]. If greenhouse gases such as $CO_2$ are the major causes of global warming, a climate sensitivity to $CO_2$ increase with a minimum error of 50\% (together with the extreme aerosol forcing uncertainty) can only raise strong doubts about the scientific robustness of the IPCC's climate change interpretation. This  error is so large because  it is not well known how to model the major climate feedback mechanisms, i.e., water vapor and clouds.\\

\begin{figure*}[tbp]
\centering
\includegraphics[width=17.0cm]{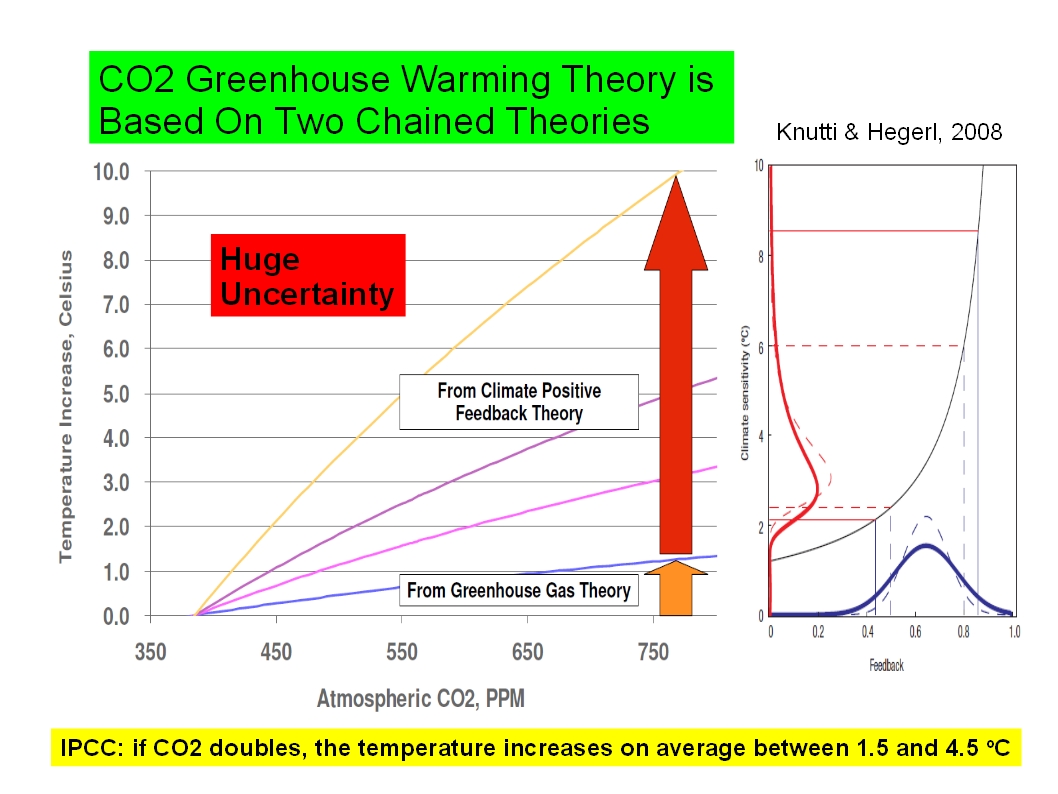}
\caption{
Climate sensitivity to  $CO_2$ doubling in function of the feedbacks (from Knutti and Hegerl [5]). Note the large uncertainty:  a $CO_2$ doubling   may cause  a global warming from  1 $^oC$ to 10 $^oC$ at equilibrium. The figure on the left explains why there exists such a large error. The GHG warming theory is based on two independent chained theories. The first theory focuses on the warming effect of a given GHG such as $CO_2$ as it can be experimentally tested. This first theory predicts that a $CO_2$ doubling    causes  a global warming of about 1 $^oC$. The second theory,  the \emph{climate positive feedback theory}, attempts to calculate the overall climatic effect of a $CO_2$ increase by assuming  an enhanced warming effect due to secondary triggering of other climatic components. For example, it is supposed that an increase of $CO_2$ causes an increase in water vapor concentration. Because $H_2O$ too is  a GHG, the overall warming induced by an increase of $CO_2$ would be due to the direct $CO_2$ warming   plus the indirect warming induced by the water vapor feedback responding to the $CO_2$ increase. The problem with the climate positive feedback theory is that it cannot be directly tested in a lab experiment. Climate modelers evaluate the climate sensitivity to  $CO_2$ increase in their \emph{climate models}, not in nature. Thus, the numerical value of this fundamental climatic component is not experimentally measured but it is theoretical evaluated with computer climate models that create virtual climate systems. It is evident that different climate models  predict a different climate sensitivity to $CO_2$, which gives rise to the huge uncertainty seen in the figure. Moreover,   if the climate models are missing important mechanisms, it is evident that  their predicted climate sensitivity to $CO_2$ changes may be extremely different from the true values.
The left figure is partially adapted from ``Catastrophe Denied: A Critique of Catastrophic Man-Made Global Warming Theory'' by  Warren Meyer,  Phoenix Climate Lecture, November 10 (2009) http://www.climate-skeptic.com/phoenix }
\end{figure*}

Indeed,  the AGWT advocates acknowledge that the current models on which the claims of the IPCC are based are significantly incomplete. Rockstrom and 28 other scientists [2], who strongly promote the AGWT, have confirmed this fact by  recently stating:
\begin{quote}
\emph{``Most models suggest that a doubling in atmospheric $CO_2$ concentration will lead to a global temperature rise of about 3 $^oC$ (with a probable uncertainty range of 2-4.5 $^oC$) once the climate has regained equilibrium. But these models do not include long-term reinforcing feedback processes that further warm the climate,... If these slow feedbacks are included, doubling $CO_2$ levels gives an eventual temperature increase of 6 $^oC$ (with a probable uncertainty range of 4-8 $^oC$).''}
\end{quote}

Rockstrom \emph{et al.} [2] gave a quite alarmist interpretation to their acknowledgment that current climate models are  missing important feedback mechanisms.
However, such alarmism is baseless.\footnote{ Rockstrom \emph{et al.} [2] have implicitly acknowledged  that the IPCC climate models  are essentially flawed. This objectively undermines the IPCC's claims because its claims  are based on the same climate models that the  AGWT advocates  acknowledge to be severely incomplete. Evidently, Rockstrom et al.'s claim that \emph{future}  climate models will necessarily confirm and greatly stress the AGWT cannot be considered as a \emph{fact}. Indeed, it cannot be ruled out that, on the contrary, \emph{future} climate models will discredit the AGWT by modeling new climate mechanisms that  current models lack.}  In fact, if missing feedback mechanisms were added to the current climate models, the corrected models would predict a much greater warming than the 0.8 $^oC$ observed during the last century.  Thus, these models would severely fail to reproduce the warming of 0.8 $^oC$ observed in the temperature data. If the current IPCC climate models do not contain many feedback mechanisms that amplify the effect of a climate radiative forcing, the logical conclusion would be that the climate sensitivity to atmospheric $CO_2$ concentration is currently significantly overestimated by those models, while the effect of the solar input is severely underestimated.\\

\section{The climatic meaning of Mann's \emph{Hockey Stick} temperature graph}

 Let us clarify the issue from a historical perspective. In 1998 and 1999 Mann \emph{et al.} [6] published the first reconstruction of global temperature over the last 1000 years. This paleoclimatic temperature reconstruction is known as the \emph{Hockey Stick} (Figure 5). This graph suggests that before 1900 the temperature of the planet was almost constant and  since 1900 an abnormal warming has occurred. From the Medieval Warm Period (1000-1300) and the Little Ice Age (1500-1750) this reconstruction predicts a cooling of less than 0.2 $^oC$. This graph surprised many, including historians and geologists who have consistently argued that the early centuries of the millennium were quite warm (the Medieval Warm Period) while the period from 1500 to 1800 was quite cold (the Little Ice Age).\footnote{This larger climate variability was clearly acknowledged  by the IPCC in 1990. It was also consistent with the world climate history after AD 1,000 according to ground borehole evidence in a paper published in 1997 (Huang S. H. N. Pollack and P. Y. Shen (1997), Late Quaternary Temperature Changes Seen in Worldwide Continental Heat Flow Measurements, \emph{Geophysical Research Letters} \textbf{24}, 1947-1950.)}\\

 The \emph{Hockey Stick} temperature graph  was considered the only global temperature reconstruction available at the time and it required a scientific interpretation. Several scientific groups, for example Crowley [7], used energy balance models and concluded that the \emph{Hockey Stick} implied that the climate is almost insensitive to solar changes and the \emph{anomalous} warming observed since 1900 has been almost entirely anthropogenic. In fact, only the ($CO_2$ and $CH_4$) GHG forcing function (as deduced from ice core reconstructions) presents  a shape that resembles that of a hockey stick. Crowley concluded his article, which shows a good correlation between his climate model and the \emph{Hockey Stick}, with this statement:
\begin{quote}
\emph{``The very good agreement between models and data in the pre-anthropogenic interval also enhances confidence in the overall ability of climate models to simulate temperature variability on the largest scales.''} (See Figure 5)
\end{quote}

\begin{figure*}[tbp]
\centering
\includegraphics[width=12.0cm]{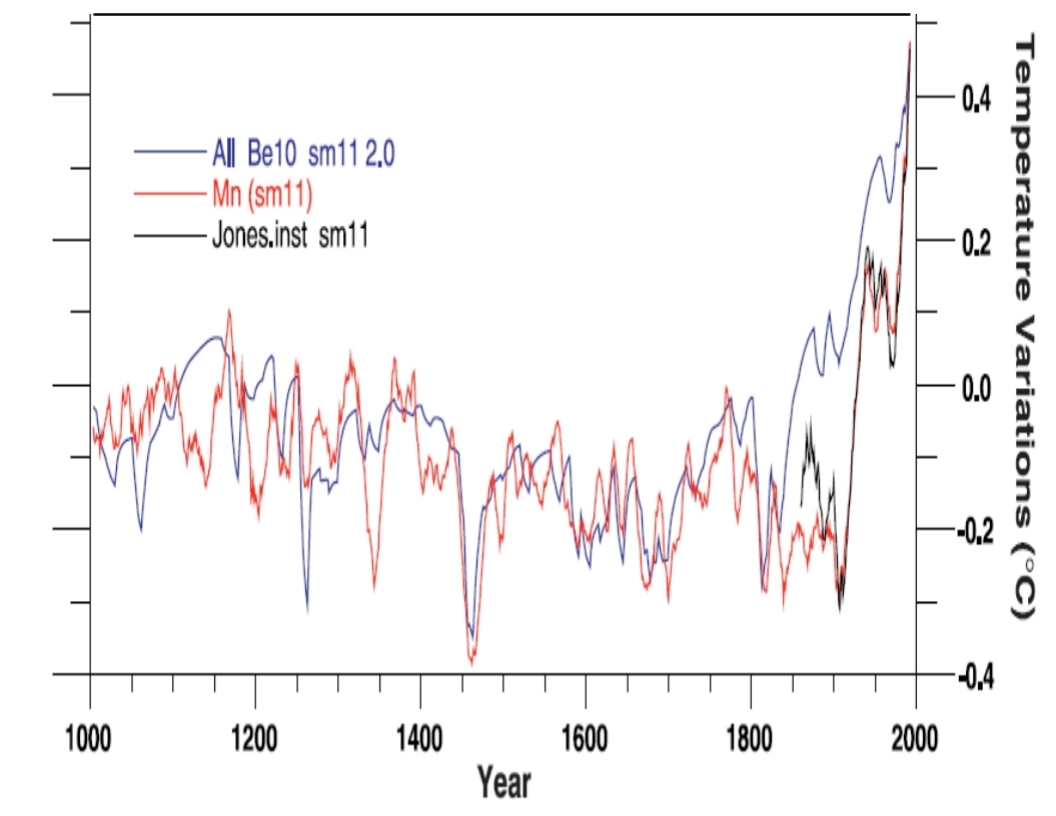}
\caption{
(Red) Mann's \emph{Hockey Stick}  [6]. (Blue) Output response of Crowley's linear upwelling/diffusion energy balance
model  using all forcing terms (solar, volcano, $CO_2$ and aerosol) [7]. Instrumental temperature data (black). Note the \emph{very good agreement} between the model and temperature reconstruction that is claimed by  Crowley in his article. Note that from the Medieval Warm Period (1000-1300) to the Little Ice Age (1500-1750) both the model response and Mann's temperature reconstruction show a cooling of about 0.2 $^oC$. }
\end{figure*}

Crowley's statement  reveals the subtle link that exists between the \emph{Hockey Stick} and the \emph{confidence} in the sufficient accuracy of the climate models used to claim that the global warming observed since 1900 was almost entirely anthropogenically induced. This interpretation was strongly endorsed by the IPCC in 2001, was popularized by Al Gore in his documentary \emph{The Inconvenient Truth}, where the \emph{Hockey Stick} plays a predominant role, and was almost completely implicitly proposed again by the IPCC in 2007. It is important to note that the IPCC's AGWT  is based on the interpretation of climate models, such as Hansen's GISS models [20], developed before 2004/5 which appear to be compatible with the \emph{Hockey Stick} temperature graph (see also Appendix H).\footnote{For example, the GISS GCM are compared and found relatively compatible with Mann's \emph{Hockey Stick} temperature graph  in Shindell D.T., G.A. Schmidt, R.L. Miller, and M.E. Mann, (2003), Volcanic and solar forcing of climate change during the preindustrial era. \emph{J. Climate} \textbf{16}, 4094-4107. A compatibility with the \emph{Hockey Stick} temperature graph is also easily visible in  the  energy balance model (EBM) simulations in Foukal P., C. Fr\"ohlich, H. Spruit, T. M. L. Wigley (2006), Variations in solar luminosity and their effect on the Earth's climate, \emph{Nature} \textbf{443}, 161-166. In this paper it is clear that the EBM model simulations predict a cooling between the Medieval Warm Period (1000-1300) and the Little Ice Age (1500-1800) of less than 0.2 $^oC$ as shown by Mann's \emph{Hockey Stick}.}\\

\begin{figure*}[tbp]
\centering
\includegraphics[width=12.0cm]{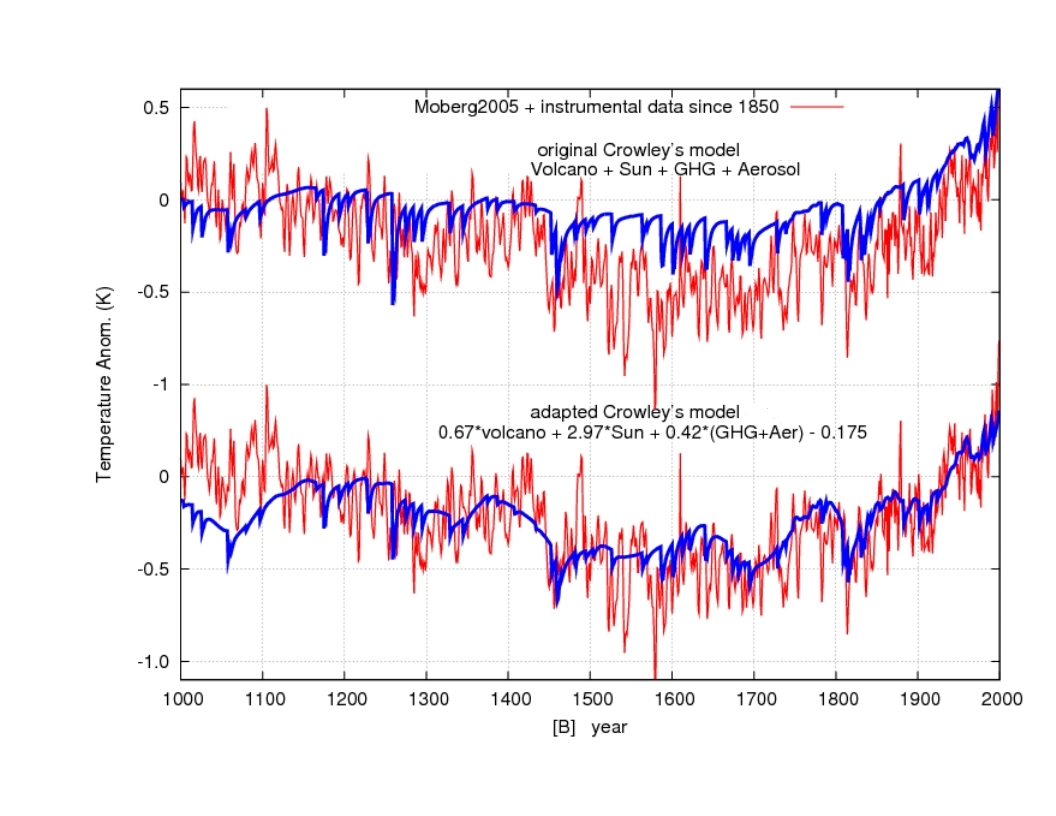}
\caption{
 Top: Moberg's temperature (red) [8]. Crowley's model (blue) [7] which is also shown in Figure 5. Bottom: In blue  Crowley's model   once adapted to reproduce the temperature of Moberg \emph{et al.} (2005) that shows a 0.6 $^oC$ cooling from MWP and LIA. Note that the solar contribution must be amplified by a  factor of 3 while the GHG+aerosol contribution, which is commonly labeled as \emph{anthropogenic}, must be reduced to a  factor of 0.4.}
\end{figure*}

\section{The climatic meaning of recent  paleoclimatic temperature reconstructions}

The dates are important because since 2004/2005 the \emph{Hockey Stick} has been mathematically and physically  questioned.\footnote{The problems with Mann's  original \emph{Hockey Stick} temperature graph  were first exposed by McIntyre  and  McKitrick (2005), The M\&M Critique of the MBH98 Northern Hemisphere Climate Index: Update and Implications, \emph{Energy and Environment} \textbf{16(1)}, 69-100. It was shown than Mann's algorithm could produce hockey stick shapes even with a set of red noise sequences. Note that recently Mann has updated his reconstruction and acknowledged that the preindustrial temperature varied more than previously claimed in his works (Mann M. E., \emph{et al.} (2008), Proxy-based reconstructions of hemispheric and global surface
temperature variations over the past two millennia, \emph{Proc Natl Acad Sci USA} \textbf{105}, 13252-13257). However, McIntyre and  McKitrick have claimed that also this latest Mann's update  presents significant mathematical errors including some records  that would be improperly used  with the axes upside down by Mann's algorithm because these records are severely compromised
by agricultural impact during the last century, e.g., Korttajarvi sediments from Tiljander data. Therefore, these data could not be used for  reconstructing the past temperature because they could not be properly calibrated against the instrumental temperature record. (McIntyre S. and R. R. McKitrick (2009) Proxy Inconsistency and Other Problems in Millennial Paleoclimate Reconstructions. \emph{Proc Natl Acad Sci USA} \emph{106}, E10.). See an extended comment by McIntyre here: http://climateaudit.org/2009/10/14/upside-side-down-mann-and-the-peerreviewedliterature/. Also interesting is the following comment by Eschenbach: http://climateaudit.org/2008/11/23/cant-see-the-signal-for-the-trees/} An additional open issue is whether the tree rings used by Mann are able to accurately reconstruct the temperature changes, especially over long time scales. Indeed,  tree growth does not depend on  temperature alone but on other factors too, such as rain patterns and biological adaptation. These multiple factors may introduce non-linear relationships and a certain degree of  randomness in the data. This may reduce the amplitude of multidecadal and secular oscillations found in the proxy models, in particular when these proxy records are statistically calibrated against the instrumental temperature records, which are only available for the period after 1850, and combined for obtaining a world average.\\

Alternative paleoclimatic reconstructions, which do not use tree rings, have been proposed [8-10]. These proxy temperature reconstructions  suggest a significant pre-industrial climate variability. From the Medieval Warm Period (1000-1300) and the Little Ice Age (1500-1750) these reconstructions show a cooling of at least 0.6 $^oC$, three times larger than the \emph{Hockey Stick}. Figure 6 shows that if Crowley's  energy balance model  is compared against Moberg's  paleoclimatic temperature reconstruction   [8],  Crowley's \emph{very good agreement} between the model and the data vanishes. If Crowley's model is recalibrated to reconstruct Moberg's temperature,   it is easy to calculate  that the solar effect must be amplified by a factor of 3 and the anthropogenic effect (GHG + Aerosol) should be multiplied by 0.4. Thus, if Moberg's  temperature  is accurate, in 2000 the anthropogenic contribution to global warming was overestimated by 250\% because of the \emph{Hockey Stick}.\\

Indeed, the \emph{Hockey Stick} temperature graph does not have any historical credibility because between 1000 and 1400, the Vikings had farms and villages on the coast of Greenland, which would suggest an even milder climate than today, while the following period, from 1400 to 1800, is known as the Little Ice Age. The medieval warm and the following cold period  were not only Western and European phenomena but are also evident in Chinese historical documents [11]. Numerous interdisciplinary studies reporting data from several regions of the world (see the Medieval Warm Period Project\footnote{\emph{http://www.$CO_2$science.org/data/mwp/mwpp.php}}) clearly indicate  a significant change in pre-industrial climate which seems to be better reproduced by more recent paleoclimatic global temperature reconstructions [8-10] which do not show a hockey stick shape.\footnote{A nice summary about the findings of numerous studies published before and after Mann's original work in 1998 and 1999 that  strongly question the validity of the \emph{Hockey Stick} temperature graph can be found  in \emph{Heaven and Earth, Global Warming the Missing Science}, chapter 2 `History,' by Ian Plimer, Taylor Trade (2009). Apparently, the  IPCC's choice in 2001 to strongly highlight  Mann's \emph{Hockey Stick} temperature graph, despite extensive published literature pointing toward a large preindustrial climate variability, can be interpreted as a further evidence of the IPCC's \emph{pro-anthropogenic} warming bias, as discussed in Section 2.} See Appendixes C-H for further details about climate data at multiple time scales.

\section{The phenomenological  solar signature since 1600}

\begin{figure*}[tbp]
\centering
\includegraphics[width=12.0cm]{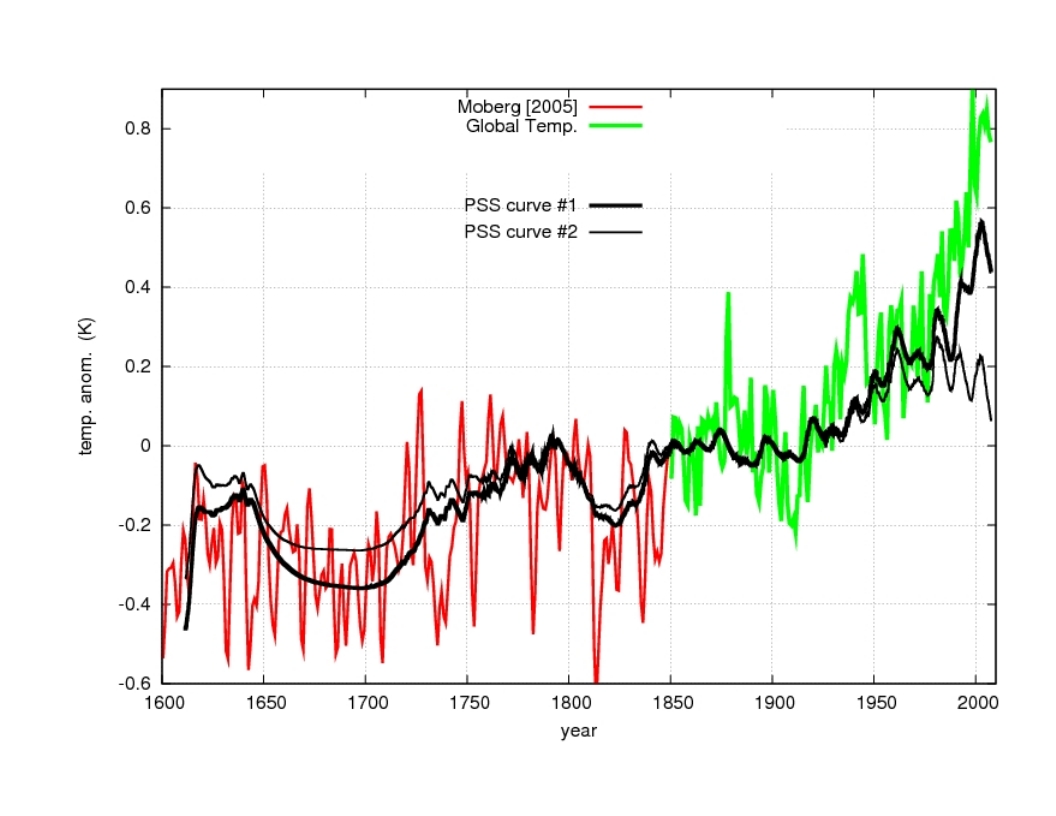}
\caption{
Reconstruction of global surface temperatures over the last 400 years (red and green). In black there is the solar signature on climate as estimated by the empirical model [12]. If since 1980 the TSI satellite composite A in figure 8  (PSS \# 1) is used most of the warming observed during the last decades is solar induced. If the TSI satellite composite C in figure 8  (PSS \# 2) is used the good correlation between temperature and solar signature  abruptly stops in 1980. (Appendix I) }
\end{figure*}

It is possible to use a phenomenological model to interpret climate change [12]. This model can simulate a typical energy balance model to interpret the global surface temperature. However, here  the climate sensitivity to solar variations and the thermodynamic characteristic relaxation times are empirically determined in the temperature patterns during the last decades.     A secular total solar irradiance reconstruction is used as input of the  model  as a proxy for the total solar activity. The model can be used to reconstruct the solar signature on  climate for the past centuries and it is possible to compare this signature against the paleoclimatic temperature reconstructions. Figure 7 shows this result:  the temperature signature induced by  solar changes as predicted by the phenomenological model well reproduces 400 years of climate change as reconstructed by Moberg \emph{et al.} [8].\\

The advantage of the phenomenological approach over that implemented in the traditional climate models, which can be described as analytic engineering, is that  the phenomenological approach   attempts to measure the climate sensitivity to solar changes through the empirical determination of a kind of \emph{response function}. This methodology would take into account all mechanisms involved in the process, although the individual microscopic mechanisms are not explicitly modeled. It is essentially analogous to the method used by an electric engineer to study the electric response properties of an unknown circuit closed inside a box by  carefully comparing the patterns of the input and the output signals.  The phenomenological approach is essentially  a holistic approach  \footnote{The term \emph{holistic science} is used as a category encompassing a number of scientific research fields. These  are multidisciplinary, are concerned with the behavior of complex systems, and recognize feedback  within systems as a crucial element for understanding their behavior. http://en.wikipedia.org/wiki/Holism\_in\_science} that emphasizes the importance of studying a complex macroscopic system by directly analyzing the properties of  the whole because of  the complex interdependence of its parts  rather than analyzing it by separating it into parts.  \\

On the contrary, the  traditional analytic climate model approach attempts to  simulate climate by dividing the climate system into its  smallest possible or discernible elements and uses their elemental physical properties alone to interpret the macroscopic system. The limitation of the latter approach is that only those mechanisms and the physical couplings among them that are currently well known can be modeled.  All unknown mechanisms and physical couplings remain excluded in an analytic model. Therefore,  the analytic modularization may fail to properly model and interpret climate change because   it just creates a \emph{virtual} climate system that may have nothing to do with reality. The risk  is \emph{scientific reductionism}, that is, compensating the current \emph{ignorance} about the true climatic mechanisms    by mistakenly stressing  a few of them, such as the anthropogenic GHG and aerosol forcings, in such a way to reproduce some warming trend during a restricted period of time. However, for not mistaking the physics of a complex phenomenon a model should be able to reproduce the data oscillations at multiple time scales. In science, the holistic approach complements the traditional analytical approach. When the two methodologies  are appropriately used  together, they are considered the most efficient way for studying complex systems. Essentially, the phenomenological approach  acknowledges that understanding climate is an \emph{inverse-problem} that risks to be \emph{ill-posed} in the analytic approach.\\

For example, the IPCC [1, p. 674] reports that the 11-year solar cycle produces a temperature cycle on the global surface temperature of about 0.1 $^oC$  that is easy to observe [12]. However, current climate models predict an average solar signature cycle which is three times smaller, approximately 0.035 $^oC$ [12] (for example, Crowley's  model [7] predicts a cycle of about 0.02 $^oC$). It is obvious that the current climate models are oversimplified. They are poorly modeling the solar-climate link mechanisms and, therefore,  mistaking the real magnitude of the solar effects on climate (Appendix J).

In fact, the IPCC  models assume that the sun can influence climate \emph{only} through  total solar irradiance variation, that is used only as a radiative forcing. However, there are additional chemical mechanisms that are stimulated by specific frequencies of the solar radiation (for example, UV alters ozone, which is a greenhouse gas, and light  stimulates  photosynthesis which influences the biosphere) and there is an additional modulation of clouds, which alters the albedo, that is due to the solar modulation of cosmic ray flux [13,14].\footnote{A solar induced low cloud cover modulation can greatly affect climate by greatly enhancing the climatic solar impact because of the potential magnitude of the resulting radiative forcing. This is evident from the fact that cloudy days are significantly cooler than sunny days. In fact, if clouds were absent the solar radiative forcing warming the Earth's surface would increase by about 30 $W/m^2$. This value is far larger than the sum of all IPCC anthropogenic forcings  in 250 years shown in Figure 2. Thus, even a small solar modulation  of  cloud cover can have a significant impact on climate change. Clouds can also respond quite fast to cosmic ray flux variations and, therefore, they may link some temperature fluctuations to the solar intermittency. Finally, a cosmic-ray cloud climate link has been suggested to explain the warm and ice periods of the Phanerozoic during the last 600 million years. In fact, the climate oscillations correlate with the cosmic ray flux variations much better than with the $CO_2$ atmospheric concentration records. In the latter case, most of the cosmic ray flux variation is claimed to be due to the changing galactic environment of the solar system, as it crosses the spiral arms of the Milky Way (Shaviv, N. J. (2003), The spiral structure of the Milky Way, cosmic rays, and ice age
epochs on Earth,  \emph{New Astronomy} \textbf{8}, 39–77;            Svensmark H. (2007), Cosmoclimatology: a new theory emerges, \emph{Astronomy \& Geophysics} \textbf{48} 1, 18-24; Kirkby J. (2009), Cosmic rays and climate, CERN Colloquium, http://indico.cern.ch/getFile.py/access?resId=0\&materialId= slides\&confId=52576)} All these alternative solar-climate link mechanisms are absent in the current climate models because the climate modelers do not know how to model them and the computers are not sufficiently fast to simulate them. The phenomenological model would automatically include all these mechanisms because the climate sensitivity to solar changes is directly, that is phenomenologically, estimated by the magnitude of the temperature patterns that can be recognized as correlated to and, therefore, likely induced by solar changes.\\

\section{The ACRIM vs. PMOD satellite total solar irradiance controversy}

Some discrepancy between the temperature reconstruction and the solar signature on climate as seen in Figures 6 and 7 may also be due to errors in the temperature as well as in the solar  proxy records.
Figure 7 shows two possible empirical solar signatures on climate after 1980.  This uncertainty is due to an uncertainty about the behavior of the total solar irradiance. The climate models adopted by the IPCC have used total solar irradiance (TSI) proxy models that claim that total solar irradiance has remained constant since 1980. However, the  satellite experimental groups (ACRIM and Nimbus7), which have measured the total solar irradiance since 1978, claim that TSI increased from 1980 to 2000 like the temperature [15].\footnote{Although it is not possible to verify the accuracy of all satellite measurements, to claim that the TSI proxy models must necessarily be correct is scientifically unsound. TSI proxy models, by definition, are based on the unproven assumptions that a given set of solar related measurements (such as sunspot number records, a few ground based spectral line width records, $^{14}C$ and $^{10}Be$ cosmogenic isotope production and others) can reconstruct TSI. However,  TSI proxy
models significantly differ from each other and, evidently, this undermines the claim that they are accurate. Thus, although TSI proxy models are useful, they cannot be used to question the accuracy of actual TSI satellite measurements without valid physical reasons. Some of the more popular TSI proxy models were produced by Lean \emph{et al.}, Solanki \emph{et al.} and Hoyt and Schatten.}

\begin{figure*}[h]
\centering
\includegraphics[width=12.0cm]{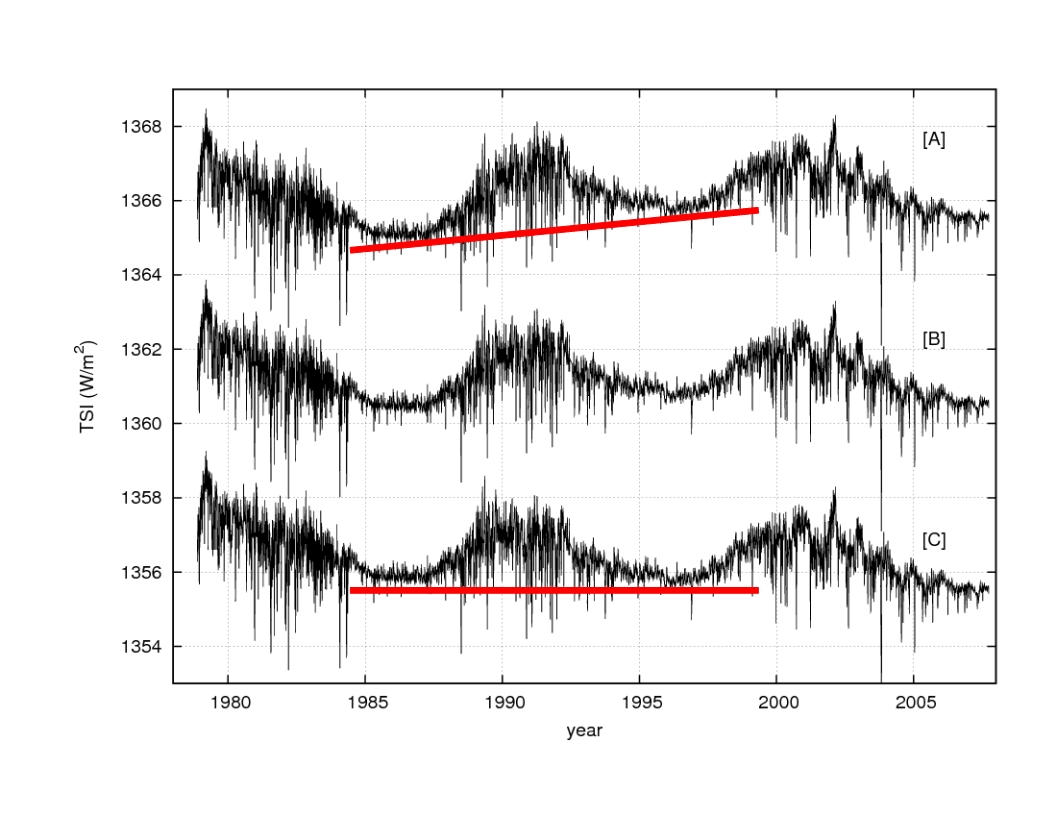}
\caption{
Possible reconstructions of the total solar irradiance using satellite data [12]. The reconstruction `A' uses the Nimbus7/ERB dataset to fill the ACRIM-gap from 1989.5 to 1991.75; the reconstruction `C' uses the ERBE/ERBs dataset to adjust the annual trend of the Nimbus7/ERB dataset and uses this altered record to fill the ACRIM-gap from 1989.5 to 1991.75; the reconstruction `C' is jus the average between the two. The level difference between the minimum in 1986 and 1996 is approximately $0.3 \pm  0.4 W/m^2$. ACRIM is between `A' and `B'; PMOD is similar to `C'. The red curves indicate the trend between the 1986 and 1996 solar minima.  }
\end{figure*}

However, another group, the PMOD in Switzerland,  claimed that  the TSI satellite data obtained and published by the above two experimental groups had to be corrected.\footnote{During the ACRIM-gap (1989.5-1992.5) Fr\"ohlich [16]  altered the Nimbus7/ERB results to make them compatible with the ERBE/ERBS results. The Nimbus7 record was shifted downward by 0.86 $W/m^2$. This shift consisted of: (1) a step function change of about 0.47 $W/m^2$ which is used to correct a hypothetical sudden change of the sensitivity of the Nimbus7's sensors following a shutdown claimed to have  occurred on 09/29/1989; (2) a linear drift of 0.142 $Wm^{-2}/yr$ from October 1989 through middle 1992 which is supposed to correct an hypothetical gradual sensitivity increase of the same satellite sensors. However, during the ACRIM-gap ERBE/ERBS sensors were expected to degrade due to a decrease in their cavity paint absorbency  which occurs during the first exposure of these kind of sensors to high solar maximum UV radiation. So, the experimental teams claim that Fr\"ohlich's alteration of the published Nimbus7/ERB data, to force them to agree with  the lower quality ERBE/ERBS results, is  unjustified.} By doing so, Fr\"ohlich obtained a TSI satellite composite  that does not show any upward trend  from 1980 to 2000 [16].
 It is important to notice that the experimental groups have always rejected the corrections of their own TSI data proposed by  PMOD  as arbitrary [15].\footnote{On September 16, 2008, Douglas Hoyt (PI of the Nimbus7/ERB experiment which is fundamental for resolving the ACRIM-gap problem, and whose data have been altered to construct the PMOD TSI satellite composite) sent me the following statement that was published in Ref. [15]: ``Concerning the supposed increase in Nimbus7 sensitivity at the end of
September 1989 and other matters as proposed by Fr\"ohlich's PMOD TSI
composite: 1. There is no known physical change in the electrically calibrated Nimbus7
radiometer or its electronics that could have caused it to become more
sensitive. At least neither Lee Kyle nor I could never imagine how such
a thing could happen and no one else has ever come up with a physical
theory for the instrument that could cause it to become more sensitive. 2. The Nimbus7 radiometer was calibrated electrically every 12 days. The
calibrations before and after the September shutdown gave no indication
of any change in the sensitivity of the radiometer. Thus, when Bob Lee
of the ERBS team originally claimed there was a change in Nimbus7
sensitivity, we examined the issue and concluded there was no internal
evidence in the Nimbus7 records to warrant the correction that he was
proposing. Since the result was a null one, no publication was thought
necessary. 3. Thus, Fr\"ohlich's PMOD TSI composite is not consistent with the
internal data or physics of the Nimbus7 cavity radiometer.
4. The correction of the Nimbus7 TSI values for 1979-1980 proposed by
Fr\"ohlich is also puzzling. The raw data was run through the same
algorithm for these early years and the subsequent years and there is no
justification for Fr\"ohlich's adjustment in my opinion.''}

\begin{figure*}[tbp]
\centering
\includegraphics[width=17.0cm]{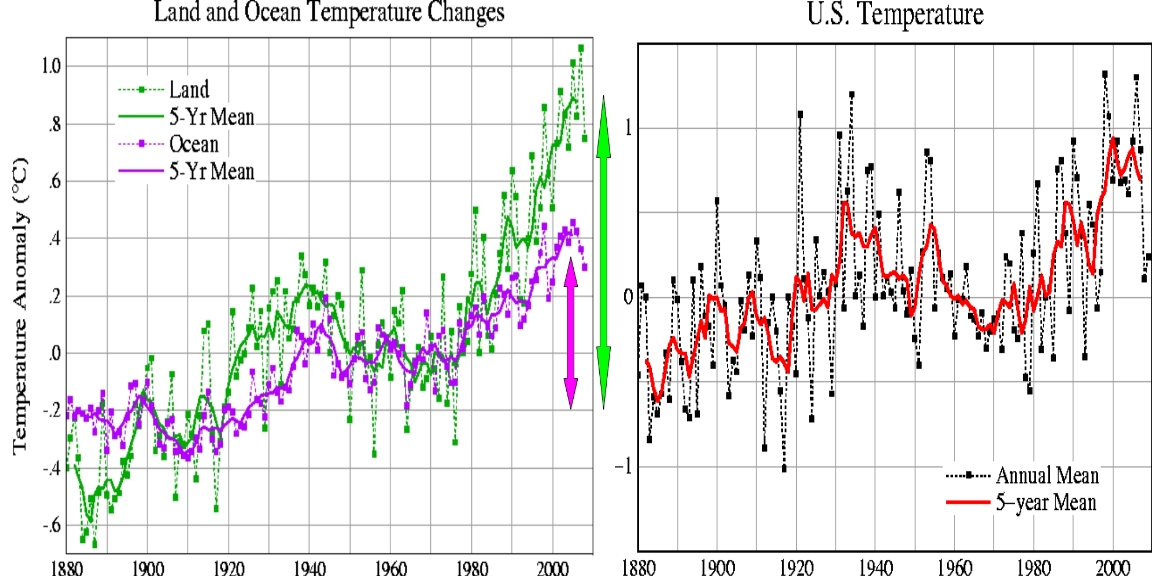}
\caption{
\textbf{Left:} Global surface temperature records of the ocean and the land. Note the significant difference observed between the two records since 1970. The land apparently warmed at a double rate than the ocean. \textbf{Right:} Global surface temperature record relative to the United States of America. Note that this record highlights the existence of a large 60-70 year cycle and shows a smaller upward secular trend.   (GISTEMP: http://data.giss.nasa.gov/gistemp/) }
\end{figure*}

By preferring the PMOD total solar irradiance satellite composite to the ACRIM one the IPCC message was that the global warming observed since 1980 could not be naturally interpreted and, therefore, it  had to be 100\% anthropogenic. Figure 8 shows three alternative reconstructions of the total solar irradiance using satellite measurements since 1978. The IPCC has adopted a reconstruction similar to C, which is compatible with PMOD's claim. However, the figure clearly indicates that the latter composite shows the lowest 1986-1996 decadal trend but, as the figure suggests, total solar irradiance could very likely have increased from 1980 to 2000 (See also Appendixes K-N).\\

\section{Problems with the global surface temperature record}

Once again it is the uncertainty in the data that makes it difficult to correctly interpret climate change. Even the global warming of about 0.8 $^oC$ since 1900 may be uncertain. In fact, during this period the land warmed by about 1.1 $^oC$, while the oceans warmed by about 0.6 $^oC$ (Figure 9). This difference appears to be too significant to be explained only by  the different thermal inertia between the ocean and the land regions. It could be partially due to an underestimation of the urban heat island effect by at least 10-20\% [17], to land use changes or perhaps to the fact that several meteorological stations located in cold regions were closed after 1960.\footnote{The surface temperature data  present several problems that may have skewed the data so as to overstate the observed warming trend both regionally and globally. For example, it has been observed that there is a significant divergence between ground temperature measurements and satellite  global temperature measurements
(Klotzbach P. J. \emph{et al.}  (2009), An alternative explanation
for differential temperature trends at the surface and in the lower troposphere, \emph{J. Geophys. Res.} \textbf{114,} D21102.)
Possible causes may be: 1) More than three-quarters of the 6,000 stations that apparently existed in the 60s were discontinued during the last decades; 2) Higher-altitude, higher-latitude, and rural locations, all of which had a tendency to be cooler, have been tendentiously removed; 3) Contamination by urbanization, changes in land use, improper siting, and inadequately-calibrated instrument upgrades further overstates warming; 4) Cherry-picking of observing sites combined with interpolation to vacant data grids may have further stressed heat-island bias; 5) Satellite temperature monitoring  findings are increasingly diverging from the station-based constructions in a manner consistent with evidence of a warm bias in the surface temperature record. For an overview on this issue see J. D'Aleo J. and A. Watts (2010), \emph{Surface Temperature Records: Policy Driven Deception?}, SPPI original paper (http://scienceandpublicpolicy.org/images/stories/papers/originals/surface\_temp.pdf)} The US temperature record present a smaller warming trend since 1880 than the global temperature records. Given the better quality of this record, this finding may suggest that part of the reported  global warming may be spurious. If the warming trend has been overestimated (or if it was partially due to land use changes), the effect of $CO_2$ and $CH_4$ on  climate change has to be reduced for this reason as well (Appendix O-P).\\

\section{A large 60 year cycle in the temperature record}

A reasonable alternative   is to extract any relevant physical information from the temperature fluctuations. It has been observed that several multi-secular climatic and oceanic records present large cycles with periods of about 50-70 years with an average of 60 years [18].\footnote{Climatic records that present a dominant cycle at about 60 year period include ice core sample, pine tree samples, sardine and anchovy sediment core samples, global
surface temperature records, atmospheric circulation index, length of the day index, etc.}  Figure 10 shows the global temperature record detrended of its quadratic upward trend [19] depicted in Figure 1. This sequence has been filtered of its fast fluctuations (by applying a six year moving average smooth algorithm) and it has been plotted against itself with a time-lag of about 60 years. The figure clearly suggests the existence of an almost perfect  cyclical correspondence between the periods 1880-1940 and 1940-2000. The peak in 1880 repeats in 1940 and again in 2000. The smaller peak in 1900 repeats in 1960.
This  60-odd year oscillation cannot be associated with any known anthropogenic  phenomenon [19]. (See also Appendixes Q and R).\\

\begin{figure*}[tbp]
\centering
\includegraphics[width=12.0cm]{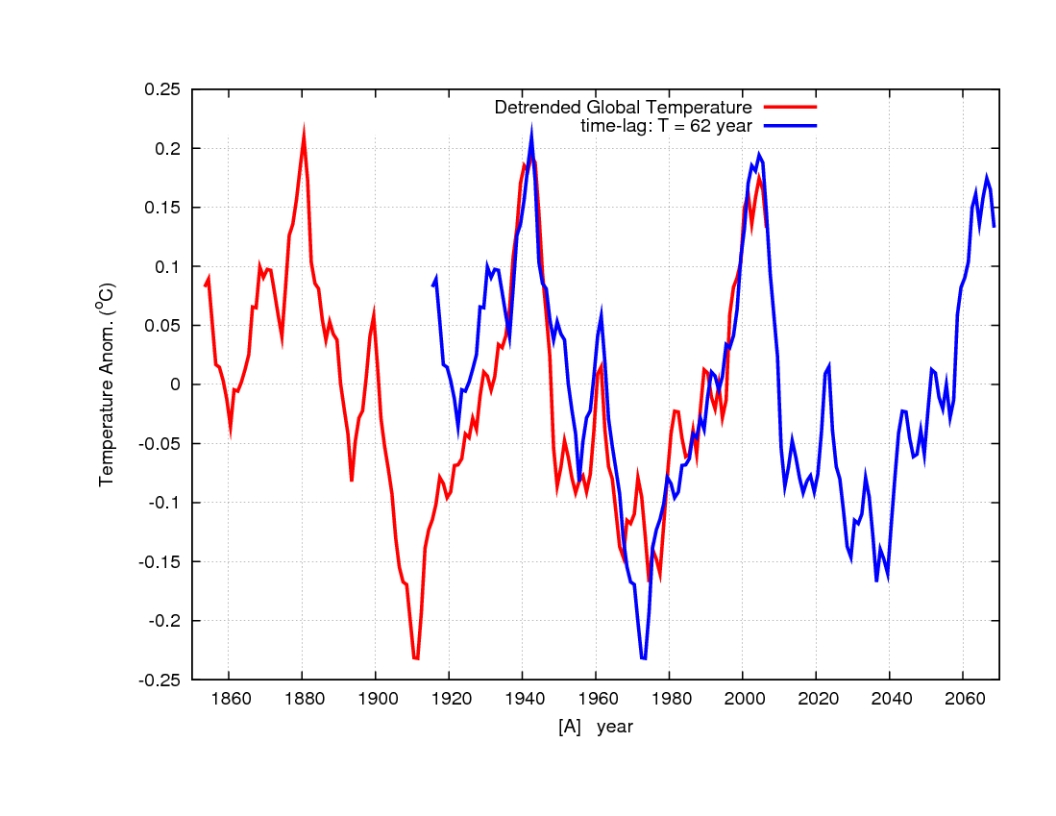}
\caption{
The 60 year cycle modulation of the temperature [19]. (Red)  global temperature detrended of its quadratic upward trend which is shown in Figure 1. (Blue) the same record time-lag shifted of about 60 years. Note the perfect symmetry between the periods 1880-1940 and 1940-2000 that excludes the fact that these cycles could have had an anthropogenic origin. Even a  smaller peak in 1900 repeats in 1960. This overwhelming clear finding, by alone,  contradicts the AGWT and the IPCC's claim that 100\% of the warming observed from 1970 to 2000 is anthropogenic. }
\end{figure*}

On the contrary, Figure 11 shows the global temperature as reproduced  by a typical climate model such as the GISS ModelE [20],  one of the major climate models adopted by the IPCC 2007. The failure of the model to reproduce the 60 year cycle is evident from the figure. Indeed, all IPCC climate models have the same  failing.\footnote{The other IPCC model scenario runs also fail to reproduce this 60-year cycle. These climate model simulations can be downloaded from the IPCC Data Archive at Lawrence Livermore National Laboratory (http://climexp.knmi.nl/selectfield\_co2.cgi?someone@somewhere). However, this is not the only shortcoming of the climate models adopted by the IPCC. These models have predicted an increase in the warming trend with altitude in the tropic troposphere due to anthropogenic GHG emissions, but balloon and satellite temperature observations have shown a significant disagreement with the model predictions. (Douglass D. H., J. R. Christy, B. D. Pearson and S. F. Singer (2007), A comparison of tropical temperature trends with model predictions,  \emph{Intl. J. Climatology}, DOI: 10.1002/joc.1651). A list of comparison of model predictions with actual observations and the incompatibility between the two was prepared by Douglas Hoyt: Greenhouse Warming Scorecard Updated 4/2/2006 (http://www.warwickhughes.com/hoyt/climate-change.htm)}\\

The existence of a natural 60 year cycle with a total (min-to-max) amplitude of at least 0.3 $^oC$, as Figure 10 shows, implies that at least 60\% of the 0.5 $^oC$ warming observed since 1970 is due to this cycle. Considering that longer natural cycle can be present and that solar activity was stronger during the second half of the 20$^{th}$ century than during the its first half [12], the natural contribution to the warming since 1970 may have been even larger than 60\%. Human emissions can have contributed at most the remaining 40\%, or less, of the warming observed since 1970 (if no overestimation of the global warming is assumed as Section 8 would suggest), not the 100\% as claimed by the IPCC.   This 60 year cycle has just entered into its cooling phase and this will likely cause a climate cooling, not a warming, until 2030-40, as Figure 10 would suggest.\footnote{There is  strong observational  evidence that the ocean has been cooling since 2003 (Loehl C. (2009), Cooling of the global ocean since 2003, \emph{Energy \& Environment} \textbf{20}, No. 1\&2, 101-104).}\\

\begin{figure*}[tbp]
\centering
\includegraphics[width=12.0cm]{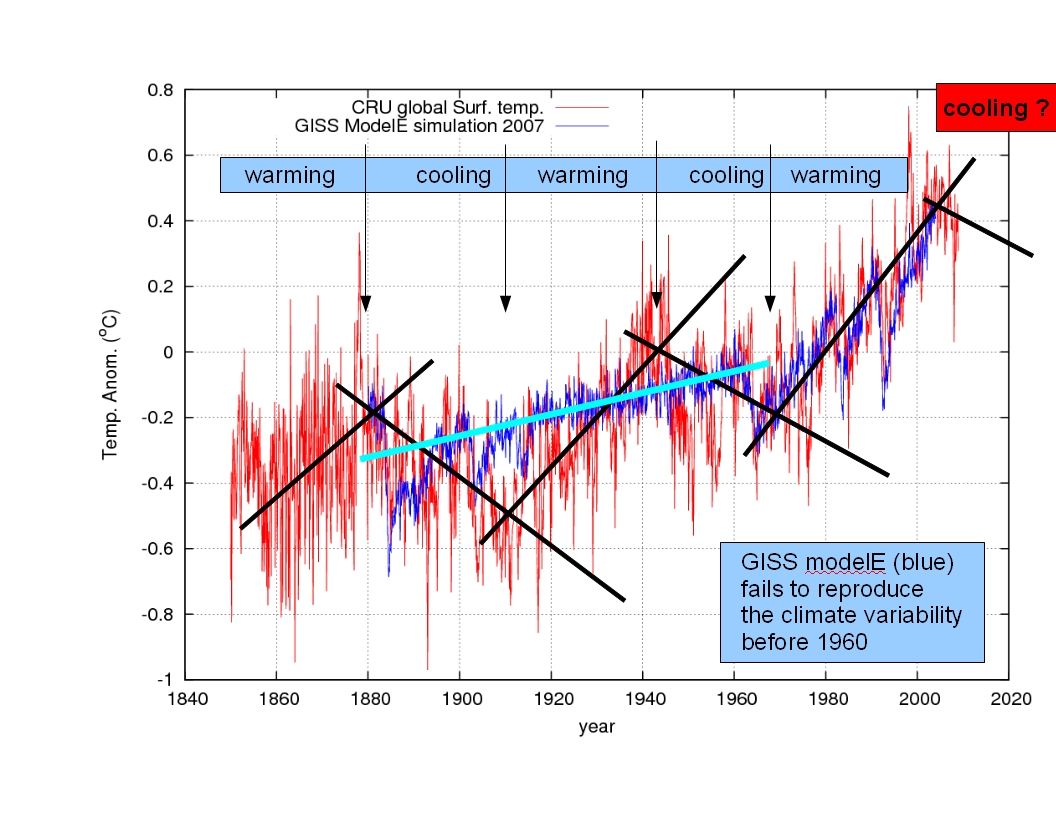}
\caption{
Global temperature (red) against the temperature prediction of the GISS ModelE [20] adopted by the IPCC. The temperature shows a clear cycle of about 60 years which herein  is emphasized by  black segments. This 60 year cycle  has been explicitly shown in Figure 10. This cycle is clearly not reproduced by the climate model simulation [19]. The climate simulation clearly crosses  from 1880 to 1970 the black segments instead of reproducing the 60 year modulation of the temperature.}
\end{figure*}

The latter result is quantitatively consistent with the results depicted in Figures 6-8 that suggest a significant change in pre-industrial climate,  in contrast  to the \emph{Hockey Stick}, and that solar activity has increased from 1980 to 2000 as Willson of the ACRIM team claims in contrast to PMOD Fr\"ohlich's  claim. In fact, they are consistent with a reduction of the anthropogenic contribution by 250\% as  calculated above in Figures 6 and 7. The independent results depicted in Figures 6, 7 and 10 are consistent with each other and would imply that if the $CO_2$ atmosphere concentration doubles,  the temperature could   rise between 1.0 and 1.5 $^oC$, which is significantly less than the IPCC's estimate of 1.5-4.5 $^oC$. \\

This result clearly indicates that  the possible impacts that anthropogenic GHGs  can have on global climate change should be greatly  diminished.  Consequently, the IPCC's claims about imminent and catastrophic  consequences that human emissions are causing and will cause, are  unsubstantiated: these claims should be greatly moderated. The existence of a large 60-year natural cycle in the global temperature  essentially  points toward the conclusion  that nature, not human activity, rules the climate.

\begin{figure*}[tbp]
\centering
\includegraphics[width=12.0cm]{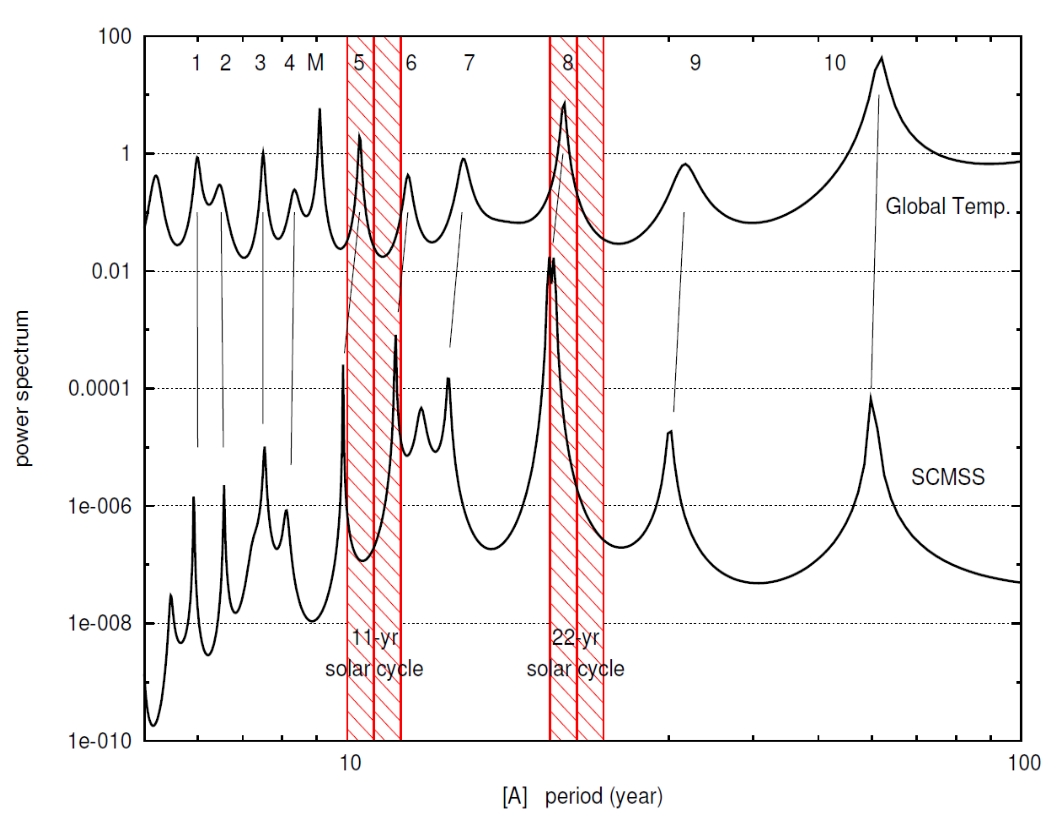}
\caption{
Spectral analysis of the global temperature from 1850 to 2009 (above), and of the speed of the center of mass of the solar system SCMSS (below) [19]. Cycles 5 and 8 are also close to the $11\pm1$ and $22\pm2$ year solar cycles. The M cycle in the spectrum of the  temperature at about 9 year is absent in the SCMSS record. However,  it corresponds to a lunar  cycle (Appendix Q-W).}
\end{figure*}

\section{Astronomical origin of the climate oscillations}

If the temperature is characterized by natural periodic cycles the only reasonable explanation is that the climate system is modulated by astronomical cycles. Natural cycles known with certainty are the  11 (Schwabe) and 22 (Hale) year solar cycles, the cycles of the planets and luni-solar nodal cycles [19].
 Jupiter has an orbital period of 11.87 years while Saturn has an orbital period of 29.4 years. These periods predict three other major cycles which are associated with Jupiter and Saturn: about 10 years, the opposition of two planets; about 20 years, their synodic cycle; and about 60 years, the repetition of the combined orbits of the two planets. The major lunar cycles are about 18.6 and 8.85 years.\\

Figure 12 shows a spectral analysis of the global surface temperature and of a record that depends on the  orbits of planets (the speed of the sun relative to the center of mass of the solar system [19]). The two records have almost the same cycles. The temperature record  contains the cycles of the planets combined with
 the two solar cycles of 11 and 22 years and a lunar cycle at about 9.1 years.\footnote{The temperature cycle `M' shown in Figure 12 appears to be exactly at 9.1 $\pm$ 0.1 years. This  period is exactly between the period of the recession of the line of lunar apsides, about 8.85 years, and  half of the period of precession of the solar-luni nodes, about 9.3 years. }  (See also Appendixes Q-V).\\

\begin{figure*}[tbp]
\centering
\includegraphics[width=12.0cm]{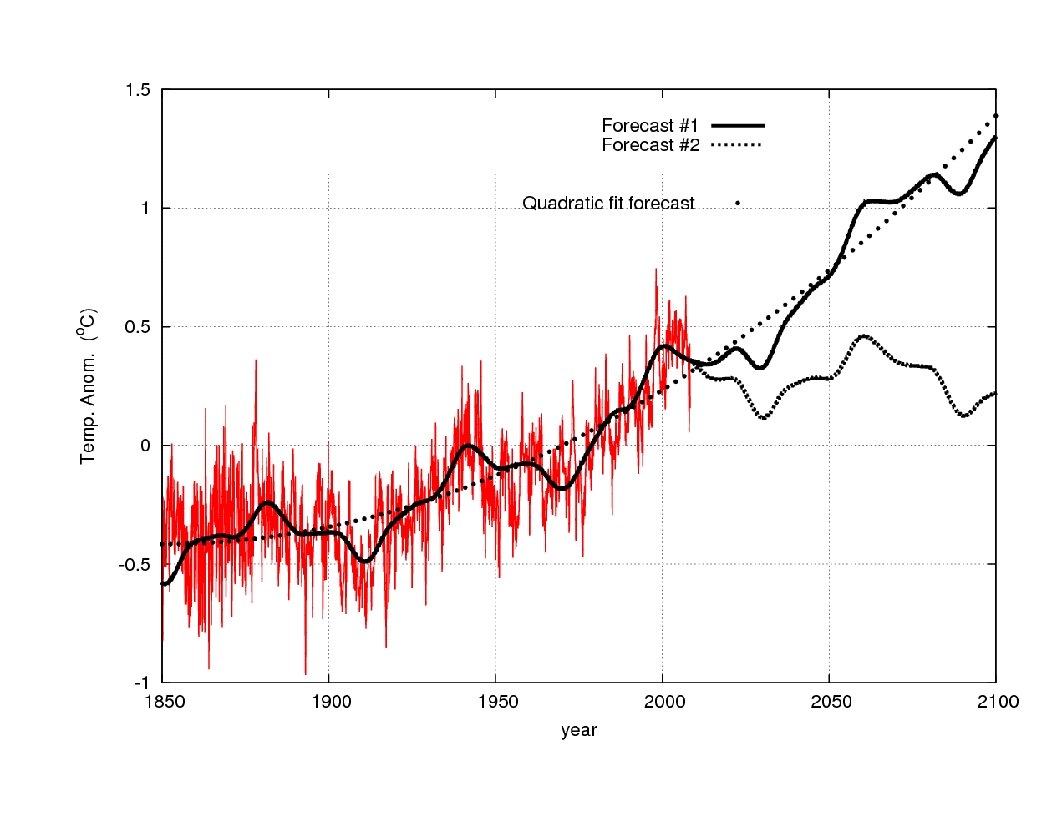}
\caption{
Global temperatures (red) and the reconstruction of temperature using only the 20 and 60 year planetary cycles (in black). The dashed curve indicates simply the quadratic trend of the temperature [19].     }
\end{figure*}

These cycles can be used to reconstruct the fluctuations of the temperature [19]. For example, it is possible to adopt a model  using only the major 20 and 60 year cycles plus a quadratic trend of the temperature and the reconstruction of Figure 13 is obtained. Other natural cycles associated with the Sun are evident in Figures 6 and 7. The model reconstructs with  great accuracy the temperature oscillations since 1850. It suggests that until  2030-2040 the temperature may remain stable if  the upward trend in temperature observed from 1850 to 2009 continues in the near future\footnote{Note that a quadratic trend function supposes a warming acceleration. Even in this situation Figure 13 would suggest that by 2100 the temperature will increase no more than 1 $^oC$ above the actual values. This estimate is significantly lower than the IPCC estimates (their figure SPM.5) that have projected a warming from 1 to 6 $^oC$ according to different GHG emission scenarios.} or the global temperature  cools if the trend of the secular solar activity decreases, as other independent considerations would suggest.\\

For example, an imminent relatively long period of low solar activity may be predicted on the basis that the latest solar cycle (cycle \#23) lasted from 1996 to 2009, and its length was about 13 years instead of the traditional 11 years. The only known solar cycle of comparable length (after the Maunder Minimum) occurred just at the beginning of the Dalton solar minimum (cycle \#4, 1784-1797)  that lasted from about 1790 to 1830. The solar Dalton minimum induced  a little ice age that lasted  30-40 years as shown in Figure 7. Therefore, it is possible that the Sun is entering into a multi-decade period of low activity, which could produce cooling of the climate. (Appendix W).\\

The physical mechanisms involved in the process are likely numerous. The gravitational forces of the planets can partially modulate the solar activity. For example, it was noted that the alignment of Venus, Earth and Jupiter presents cycles of approximately 11 years that are in phase with the 11-year solar cycles [21] and multi secular reconstructions of solar activity reveal 60-year cycles associated with the combined orbit of Jupiter and Saturn and other longer cycles [22]. Solar changes could modulate climate change through various physical and chemical processes as explained in Section 6, which are currently not included in the models, as explained in Section 6.\\

\begin{figure*}[tbp]
\centering
\includegraphics[width=12.0cm]{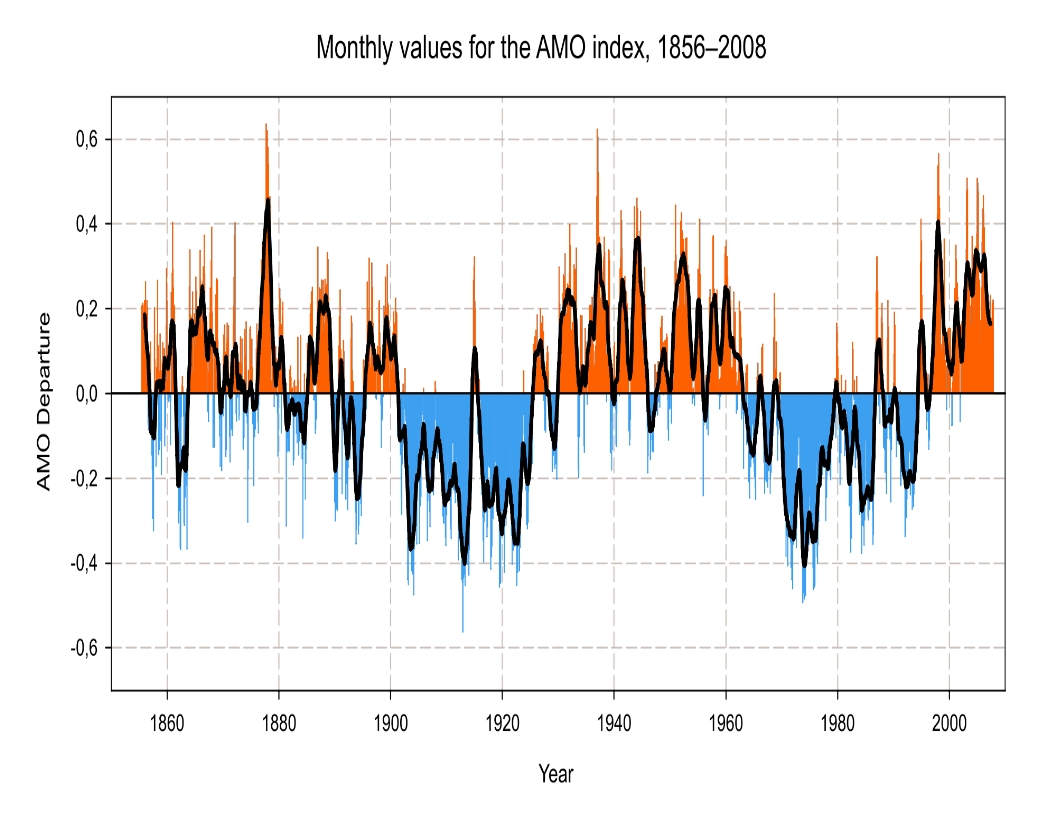}
\caption{
Monthly value of the Atlantic Multidecadal Oscillation (AMO) index. Note the evident 60-year cycle.       The figure is adapted from Wikipedia at the voices:  \emph{Atlantic Multidecadal Oscillation}.}
\end{figure*}

 There is also the possibility that the Earth's orbital parameters are directly modulated by the gravitational forces of Jupiter, Saturn and the Moon, and the Sun's magnetic force in such a way that the length of day is modulated and/or other planetary parameters are altered. For example, the rotation of the Earth on its axis shows 60-year cycles that  anticipate those of the temperature by a few years [18, 23]. Variations in the Earth's rotation and tides caused by the lunar  cycles can drive  ocean oscillations, which in turn may alter the  climate [19]. For example,   the Atlantic Multidecadal Oscillation (AMO) and the  Pacific Decadal Oscillation (PDO) present clear 60-year cycles and other faster cycles, see Figures 14 and 15. None of these mechanisms are included  in the models adopted by the IPCC.\\

\begin{figure*}[tbp]
\centering
\includegraphics[width=18.0cm]{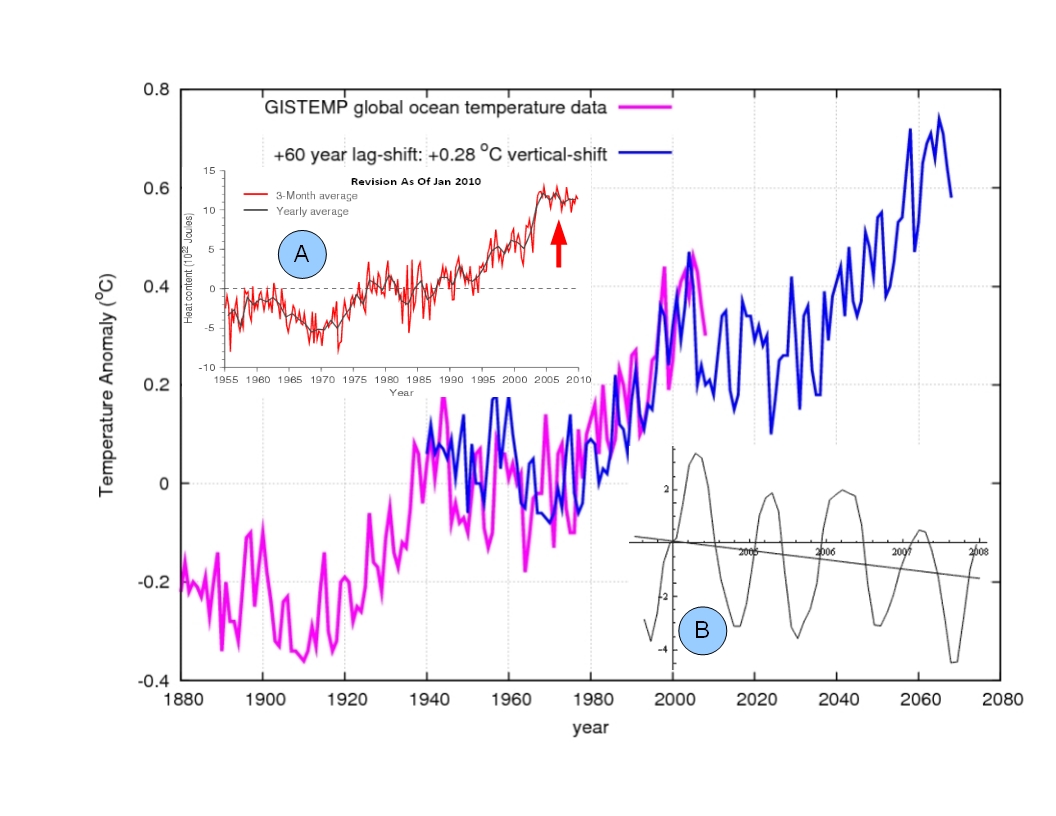}
\caption{
GISTEMP global surface ocean temperature record (shown also in Figure 9) plotted against itself with a 60 year lag shift and a vertical translation. The figure clearly indicates an almost perfect matching between the 1880-1940 and 1940-2000 periods. This further proves the existence of a 60-year natural cycle driving climate change. In particular, note the almost perfect correspondence between the warming trends during the 1910-1940 and 1970-2000 periods. The upward trend during the former period (1910-1940) appears even slightly higher than the upward trend during the latter period (1970-2000) suggesting a warming deceleration.  This finding strongly contradicts the IPCC's claim that 100\% of the warming observed since 1970 can only be explained with anthropogenic emissions. The figure also suggests a slight ocean cooling until 2030-2040, as does the climate-planetary model of Figure 13,  which started  in 2002-2003. The upward trend between the first and second half of the last century correlates well with the increased solar activity occurring between the first and second half of the 20$^{th}$ century, as indicated by all solar models (see Ref. [12]  and Figure 7). If the upward warming trend observed since 1880 continues the ocean will not likely warm more than 0.5 $^oC$ by 2100.\newline
\textbf{The insert A} shows the latest update (Jan/2010) of the Ocean Heat Content (OHC) data from the National Oceanographic Data Center (NODC) (http://www.nodc.noaa.gov/OC5/3M\_HEAT\_CONTENT/). This record shows a decrease of the ocean heat content since 2003. This record significantly differs from an earlier version, which showed a slight increase, published in Levitus \emph{et al.} (2009), Global ocean heat content 1955-2008 in light of
recently revealed instrumentation problems, \emph{Geophys. Res. Lett.} \textbf{36}, L07608.  \newline \textbf{The insert B}  depicts ocean heat content as measured by Argo. Argo is a network of over 3000 floats scattered across the globe that measure temperature and salinity of the upper ocean. A cooling trend from 2003 to 2008 is found.  Loehl C. (2009), Cooling of the global ocean since 2003, \emph{Energy \& Environment} \textbf{20}, No. 1\&2, 101-104.\newline
Both inserts are compatible with a forecast suggesting that ocean temperature will decrease until 2030-2040. }
\end{figure*}

\section{Conclusion}

The analysis of several records  suggests that the IPCC's claim that humanity is running an imminent danger because of anthropogenic $CO_2$ emissions\footnote{The AGWT advocates  claim, by using climate model projections, that an increase in anthropogenic  $CO_2$ concentration in the atmosphere will lead to ecological disasters, including wild swings in weather patterns, extended desertification, spread of hot-climate infectious diseases,  greater risks of severe damaging weather phenomena such as Katrina-like hurricanes, melting of the glaciers in a few decades that, in turn, will leave hundreds of  millions of people without fresh water, cause the extinction of polar bears and raise so much the ocean level that  all coasts and  their cities will be severely flood beginning, of course, with New York [Al Gore (2006), \emph{An Inconvenient Truth,}  documentary movie].   After that, an increase in anthropogenic  $CO_2$ will reach the \emph{tipping} point and activate a runaway greenhouse effect that will let the oceans boil away and, ultimately, transform the Earth into a Venus-like state (as James Hansen claimed during his AGU 2008 scientific talk (2008-12-17) ``Climate Threat to the Planet''). All this AGWT apocalypticism is extensively rebutted in \emph{Climate Change Reconsidered} [3] by using scientific research based on actual data. Al Gore's movie has been elegantly rebutted by Christopher Monckton of Brenchley in \emph{``35 Inconvenient Truths:
The errors in Al Gore's movie''} SPPI (2007),  http://scienceandpublicpolicy.org/monckton/goreerrors.html \\
Regarding the hypothetical \emph{tipping} point and the runaway greenhouse effect that the Earth would be risking,  it should be noted that the atmospheric  $CO_2$ concentration was many times higher than today in almost all earlier geologic periods when  no runaway greenhouse effect occurred (Hayden H. C. (2007), \emph{A Primer on $CO_2$ and Climate}, Vales Lake Publishing, LLC.). For example, during the Jurassic period (150-200 million years ago) the $CO_2$ concentration was at least 5 times higher than today (about 2000 ppmv), and  during the Cambrian period (500-550 million years ago) it was at least 10-15 times higher than today (4000-6000 ppmv). Interestingly, during the late Ordovician period (490-440 million years ago) the Earth experienced an extremely cold glacial period despite the fact that the $CO_2$ concentration was at least 10 times higher than today. In fact,  most of the greenhouse effect that keeps the Earth  warm is regulated by water vapor, not by $CO_2$.  The water vapor concentration, together with the low cloud cover percentage, are not well understood yet but they are likely strongly influenced by the solar changes and cosmic rays. During the last few years there has been  a tendency among the AGWT advocates to declare $CO_2$ to be a \emph{pollutant}. This is a further serious mystification of the reality. For human health $CO_2$ is completely innocuous even at a concentration 10 times larger that the 0.039\% (390 ppmv) actual atmospheric value. Moreover,  $CO_2$ is as essential to life as oxygen and water. Carbon dioxide is the major food for  plants, which in turn  are food for animals, and of course for humans too. Indeed, an increase in atmospheric $CO_2$  concentration  would lead to accelerated  plant growth and, therefore, to increased food production [3]. In fact, in man-made greenhouses $CO_2$ is enriched  at 2, 3 or 4 times the natural concentration (about 1000 ppmv) because  this causes plants to grow faster and improves plant quality. Thus, an increase of atmospheric $CO_2$  concentration  may also benefit humanity.}  is based on  climate models that are too simplistic. In fact, these models  fail to reproduce the temperature patterns and the temperature oscillations at multiple time scales. (See also Appendixes H, J, X-Z). These models exclude several mechanisms that are likely to affect  climate change related to natural temperature oscillations that have nothing to do with man.  Indeed, these oscillations, such as a large 60 year cycle,  appear to be synchronized with the oscillations of the solar system.\\

 By ignoring these natural mechanisms, the IPCC, also through a questionable choice of data and labels as explained in Section 2, has greatly overestimated the effect of an anthropogenic forcing by a factor between  2 and 3 just to fit the observed global warming in particular from 1970 to 2000, as the climate model depicted in Figure 11 shows. However, a detailed climatic reconstruction suggests that the phenomenological model depicted in Figures 13 and 15 is more satisfactory and is likely to be more accurate in forecasting  climate change during the next few decades, over which time  the global surface temperature  will likely remain steady or actually cool.\footnote{Some of the strongest AGWT advocates are rapidly acknowledging that no significant global warming has been observed since at last 10 years contrary to the IPCC  projections. Other factors, besides anthropogenic GHGs, are responsible of climate changes. In February 2010 Phil Jones, the ex director of the CRU center for climate change and the academic at the center of the \emph{climategate},
  has admitted that there has been no global warming since 1995.
http://www.dailymail.co.uk/news/article-1250872/Climategate-U-turn-Astonishment-scientist-centre-global-warming-email-row-admits-data-organised.html \\
   In 2009 Susan Solomon \emph{et al.} (Irreversible climate change due to carbon
dioxide emissions, PNAS \textbf{106} 1704-1709) predicted  a large, imminent and irreversible warming since 2000 due to anthropogenic emissions. However, in January 2010  Solomon \emph{et al.} (Contributions of stratospheric water vapor to decadal changes in the rate of global warming, \emph{Science} 10.1126/science.1182488) acknowledged that stratospheric water vapor, not just anthropogenic $CO_2$ and $CH_4$,
is an important climate driver of decadal global surface
climate change
 that has largely contribute both to the warming observed from 1980-2000 (30\%) and to the slight cooling observed after 2000 (25\%). Stratospheric water vapor concentration can also be indirectly driven by UV solar irradiance variations through ozone modulation and its contribution would be included in the phenomenological model herein presented.}

}

\end{document}